# Nonlinear Modal Decoupling of Multi-Oscillator Systems with Applications to Power Systems

Bin Wang, *Student Member, IEEE*, Kai Sun, *Senior Member, IEEE*, and Wei Kang, *Fellow, IEEE*

*Abstract*—Many natural and manmade dynamical systems that are modeled as large nonlinear multi-oscillator systems like power systems are hard to analyze. For such a system, we propose a nonlinear modal decoupling (NMD) approach inversely constructing as many decoupled nonlinear oscillators as the system's oscillation modes so that individual decoupled oscillators can easily be analyzed to infer dynamics and stability of the original system. The NMD follows a similar idea to the normal form except that we eliminate inter-modal terms but allow intra-modal terms of desired nonlinearities in decoupled systems, so decoupled systems can flexibly be shaped into desired forms of nonlinear oscillators. The NMB is then applied to power systems towards two types of nonlinear oscillators, i.e. the single-machine-infinite-bus (SMIB) systems and a proposed non-SMIB oscillator. Numerical studies on a 3-machine 9-bus system and New England 10-machine 39-bus system show that (i) decoupled oscillators keep a majority of the original system's modal nonlinearities and the NMB provides a bigger validity region than the normal form, and (ii) decoupled non-SMIB oscillators may keep more authentic dynamics of the original system than decoupled SMIB systems.

*Index Terms*— Nonlinear modal decoupling, inter-modal terms, intra-modal terms, oscillator systems, normal form, power systems, nonlinear dynamics.

## I. INTRODUCTION

Oscillator systems, i.e. a system with a number of oscillators interacting with each other, are ubiquitous in both natural systems and manmade systems. In biological systems, low-frequency oscillations in metabolic processes can be observed at intracellular, tissue and entire organism levels and they have a deterministic nonlinear causality [1]. In electric power grids, which are among the largest manmade physical networks, oscillations are continuously presented during both normal operating conditions and disturbed conditions [2]. In some fields of both natural science and social science, the Kuramoto model is built based upon a large set of coupled oscillators modeling periodic, self-oscillating phenomena in, e.g., reaction-diffusion systems in ecology [3] and opinion formation in sociphysics [4]. For all these oscillator systems, the common underlying mathematical model is actually a set of interactive governing differential equations, linear or nonlinear.

An ideal way to study dynamics of a multi-oscillator system from an initial state is to find an analytical solution of its differential equation models and use the solution for further prediction and control. However, even finding an approximate solution of a high-dimensional nonlinear multi-oscillator system has been a challenge for a long time to mathematicians, physicists and engineers [5]. Analytical efforts have been made in broader topics, like dynamical systems [6][7], nonlinear oscillations [8] and complex networks [9], to better understand, predict and even control the oscillator systems, and some well-known theories are such as the perturbation theory and Kolmogorov-Arnold-Moser theory. Most of these efforts attempt to directly analyze an oscillator system as a whole and extract desired information, e.g. approximate solutions and stability criteria, from the governing differential equations. Especially, extensive attentions recently have been paid to using the theory of synchronization to analyze the interactions among oscillators in a system [10]-[14]. In addition, numerical studies can provide dynamical behaviors of high-dimensional oscillation systems with desired accuracy. However, simulating a high-dimensional oscillator system like a power grid could be very slow if oscillators are coupled through a complex network and interact nonlinearly [15].

In this paper, we aim at *inversely constructing* a set of decoupled, independent oscillators from a given high-dimensional multi-oscillator system. Each of those decoupled oscillators is a fictitious $2^{nd}$ order nonlinear system that corresponds to a single oscillation mode of the original system. For some real-life oscillator networks such as a power grid networking synchronous generators, those real oscillators themselves often have strong couplings and interactions in dynamics. However, the modal dynamics with respect to different oscillation modes may have relatively weak couplings or interferences unless significant resonances happen between oscillation modes. Thus, the fictitious oscillators that are inversely constructed to represent different oscillation modes are independent, or in other words naturally decoupled, to some extend and hence can be more easily understood and analyzed to gain insights on the dynamical behaviors, stability and control of the whole original system. In this paper, we define such a process as *nonlinear modal decoupling*, i.e. inverse

This work was supported by NSF CAREER Award (ECCS-1553863).
B. Wang and K. Sun are with University of Tennessee, Knoxville, TN 37996 USA. (e-mail: {bwang, kaisun}@utk.edu).
W. Kang is with the Department of Applied Mathematics, Naval Postgraduate School, Monterey, CA 93943 USA. (e-mail: wkang@nps.edu).

construction from the original nonlinear multi-oscillator system to a set of decoupled fictitious nonlinear oscillators.

Finding the modal decoupling transformation even for general linear dynamical systems has been studied for more than two hundred years, and massive papers aimed at decoupling linear dynamical systems with non-classical damping. In the 1960s, Caughey and O'Kelly [16] found the necessary and sufficient conditions for a set of damped second-order linear differential equations to be transformed into decoupled linear differential equations based on early mathematical works by Weirestras in the 1850s [17]. In just the last decade, the decoupling of linear dynamical systems with non-classical damping was achieved [18]-[20].

For nonlinear oscillator systems, the modal decoupling has not been studied well despite its importance in simplification of stability analysis and control on such complex systems. Some related efforts have focused on transformation of a given nonlinear oscillator system towards an equivalent linear system . One approach is feedback linearization that introduces additional controllers to decouple the relationship between and outputs and inputs in order to control one or some specific outputs of an oscillator system [21]-[24]. Another approach is normal form [25]-[29] that applies a series of coordinate transformations to eliminate nonlinear terms starting from the $2^{nd}$ order until the simplest possible form. If regardless of resonances, such a simplest form is usually approximated by a linear oscillator system, whose explicit solution together with the involved series of transformations are then used to study the behavior of the original nonlinear oscillator system. To summarize, to analyze the high dimensional nonlinear oscillator systems, the efforts in the present literature tend to achieve an approximate or equivalent linear system so as to utilize available linear analysis methods. Based on these efforts, it is quite intuitive to move one step forward to achieve a set of decoupled nonlinear oscillator systems where each are simple enough for analysis of dynamics and stability. That is the objective of this work.

For real-life nonlinear oscillator systems such as a multi-machine power system, a linear decoupling, if exists, can transform the system into its modal space, which may help improve the modal estimation [30] and assess the transient stability of the system [31]. The normal form method was introduced to power systems in 1992 by paper [32] for analyzing stressed power systems and enables the design of controllers considering partial nonlinearities of the systems. Since nonlinearities are considered, like the $2^{nd}$ order nonlinearity in [33] and the $3^{rd}$ order nonlinearity in [34], the approximated solution from normal form may have a larger validity region than the linearized system [35]. Among these attempts, including [31] on a nonlinear modal decoupling, the results do not provide the corresponding nonlinear transformation that links the original oscillator system to the nonlinear modal decoupled systems.

In this paper, the proposed nonlinear modal decoupling approach is derived adopting an idea similar to the Poincaré normal form in generating a set of nonlinear homogeneous polynomial transformations [36]. However, different from the classic theory of normal forms, we *eliminate only the inter-modal terms and allow decoupled systems to have intra-modal terms of desired nonlinearities for nonlinear modal decoupling*.

The rest of the paper is organized as follows: in Section II, the definitions, derivations and error estimation indices on the nonlinear modal decoupling are presented. In Section III, the nonlinear modal decoupling approach is applied to multi-machine power systems and its application in first-integral based stability analysis. Section IV shows numerical studies on the IEEE 3-machine 9-bus power system and IEEE 10-machine 39-bus power system. Conclusions are drawn in Section V.

## II. NONLINEAR MODAL DECOUPLING

We will first introduce several definitions and one lemma before presenting Theorem 1 on nonlinear modal decoupling.

Given a nonlinear dynamical system described by a set of ordinary differential equations:

$$\dot{X} = F(X) \qquad (1)$$

where $X$ is the vector containing $N$ state variables and $F$ is a smooth vector field. The origin is assumed to be an equilibrium point (if not, it can be easily moved to the origin by a coordinate transformation)

In this paper, the given dynamical system in (1) is called a *multi-oscillator system* if and only if all eigenvalues of its Jacobian matrix, say $A$, appear as conjugate pairs of complex numbers. Each conjugate pair defines a unique *mode* of the system. Let $\Lambda = \{\lambda_1, \lambda_2,..., \lambda_N\}$ represent the matrix of $A$'s eigenvalues, where $N$ is an even number. Without loss of generosity, let $\lambda_{2i-1}$ and $\lambda_{2i}$ be conjugate pair corresponding to the mode $i$.

**Definition 1 (Desired modal nonlinearity)** If the multi-oscillator system (1) can mathematically be transformed into the form (2) and the two governing differential equations in (2) regarding mode $i$ have $\mu$-coefficients of desired values, then the $i$-th mode is said to have the desired modal nonlinearity.

$$\dot{z}_{2i-1} = \lambda_{2i-1}z_{2i-1} + \sum_{\alpha=1}^{N}\sum_{\beta=\alpha}^{N}\mu_{2i-1,\alpha\beta}z_{\alpha}z_{\beta} + \cdots$$
$$+ \sum_{\alpha=1}^{N}\sum_{\beta=\alpha}^{N}\cdots\sum_{\rho=\gamma}^{N}\mu_{2i-1,\alpha\beta\cdots\rho}\underbrace{z_{\alpha}z_{\beta}\cdots z_{\rho}}_{k \text{ terms in total}} + \cdots \qquad (2)$$
$$\dot{z}_{2i} = \bar{\dot{z}}_{2i-1}$$

where $Z=\{z_1, ..., z_N\}$ is the vector of state variables and $k > 1$.

In the traditional normal form method, only the modal nonlinearities that cannot be eliminated due to resonance are retained, which is equivalent to making as many $\mu$-coefficients be zero as possible in (2). If regardless of the resonance, the advantage of the standard normal form is that the resulting truncated system will be a linear dynamical system having an analytical solution.

However, it is not always true that a linear system is the most desired. For instance, in power systems, power engineers and researchers prefer to assume that the underlying low-dimensional system dominating each nonlinear oscillatory mode follows the nonlinearity of a single-machine-infinite-bus (SMIB) power system [31][46][47], i.e. the simplest single-degree-of-freedom power system. Thus, this paper is motivated to keep specific nonlinear terms for the desired modal nonlinearity by following either the SMIB assumption, as shown in Section III-B, or another assumption proposed in Section III-C.

For the normal form method, the truncated linear system cannot be used for estimating the boundary of stability, which is meaningful for a nonlinear system. As a comparison, the nonlinear modal decoupling to be proposed provides the possibility to estimate the boundary of stability using the nonlinearities intentionally kept in the model, although estimation of the stability boundary of a truncated system model is a long standing problem. An approximation of the stability boundary will be presented in Section III-D.

For the convenience of statements, the following definitions are adopted to introduce which nonlinear terms should be kept or eliminated.

**Definition 2 (Intra-modal term and inter-modal term)** Given the desired modal nonlinearity (2) for mode $i$ of the multi-oscillator system (1), the *intra-modal terms* are the nonlinear terms in the form of $\mu_{j,\alpha\beta\cdots\rho}z_\alpha z_\beta\cdots z_\rho$ (for $k = 2, 3, \cdots$) which involve state variable(s) only corresponding to mode $i$, i.e. indices $j,\alpha,\beta,\cdots,\rho \in \{2i-1, 2i\}$. All the other nonlinear terms are called the *inter-modal terms*, which involve state variables corresponding to other modes.

**Definition 3 (Mode-decoupled system)** If the form (2) with desired modal nonlinearity regarding the $i$-th mode also makes (3) satisfied, then (2) is called *a mode-decoupled system* for mode $i$.

$$\mu_{j,\alpha\beta\cdots} = \begin{cases} \text{desired value} & \text{if } j,\alpha,\beta,\cdots, \in \{2i-1, 2i\} \\ 0 & \text{otherwise} \end{cases} \quad (3)$$

Based on the concept of resonance [36], the $n$-triple $\Lambda = \{\lambda_1, \lambda_2, \ldots, \lambda_N\}$ of eigenvalues is said to be *resonant* if among the eigenvalues there exists an integral relation $\lambda_s = \sum_k m_k\lambda_k$, where $s$ and $k = 1, \ldots, N$, $m_k \geq 0$ are integers and $\sum_k m_k \geq 2$. Such a relation is called a *resonance*. The number $\sum_k m_k$ is called the *order of the resonance*. Now, we present the Theorem on nonlinear modal decoupling.

**Theorem 1 (Nonlinear modal decoupling)** Given a multi-oscillator system in (1) and a desired modal nonlinearity having inter-modal terms eliminated, if resonance does not exist for any order, then (1) can be transformed into (2) by a certain nonlinear transformation, denoted as $H$.

**Remark** The rest of this section will focus on giving a constructive proof of the theorem, in which we derive the transformation $H$ and its inverse that can be numerically computed. This is different from the normal form theory where the focus is on the existence of the normal form. Unlike the normal form, the nonlinear modal decoupling requires elimination of only inter-modal terms so as to decouple the dynamics regarding different modes while leaving room for intra-modal terms to be designed for desired characteristics with each mode-decoupled system. For simplicity, we use $\mu_{\text{intra}}$ and $\mu_{\text{inter}}$ to respectively call the intra- and inter-modal term coefficients. For this section, we assume that the desired modal nonlinearity for each mode to be known. Nonlinear modal decoupling on a real-life high-dimensional multi-oscillator system like a power system might intentionally make each mode-decoupled system have the same modal nonlinearity as a single-oscillator system of the same type, e.g. an SMIB system for power systems, for the convenience of using the existing methods on the same type of systems. However, for the purpose of stability analysis and control, decoupling a real-life system into a different type of oscillators might also be an option. In the next section, two ways to choose the desired modal nonlinearity will be illustrated on power systems.

The detailed derivation of the transformation $H$ used for nonlinear modal decoupling will be presented in the constructive proof of Theorem 1, where $H$ will be a composition of a sequence of transformations, denoted as $H_1, H_2, \ldots, H_k, \ldots$, where $H_k$ is a homogeneous polynomial. The relationship between the state variables of the mode-decoupled system, say $Z$, and the state variables after the $k$-th transformation are shown in (4) based on $H_1, H_2, \ldots, H_k, \ldots$, where we use the "$Z^{(k)}$" to represent the vector of state variables after the $k$-th transformation.

$$X = H(Z) = \cdots = (H_1 \circ H_2 \circ \cdots \circ H_k \circ \cdots)(Z)$$
$$Z^{(1)} = (H_2 \circ \cdots \circ H_k \circ \cdots)(Z)$$
$$\cdots \qquad\qquad\qquad\qquad\qquad\qquad\qquad (4)$$
$$Z^{(k)} = (H_{k+1} \circ \cdots)(Z)$$
$$\cdots$$

We first introduce a lemma before presenting the proof of Theorem 1.

**Lemma 1.** Given one transformed form (5) of a multi-oscillator system, where $D_j$ only contains intra-modal terms and $C_j$ only contains inter-modal terms and they are vector polynomial functions of degree $j$ in $Z^{(p)}$. If resonance does not exist up to the order $p+1$, then in a certain neighborhood of the origin of $Z^{(p+1)}$, denoted as $\Omega_{p+1}$, the inter-modal terms of degree $p+1$ can be completely eliminated to give (6) by a polynomial transformation of degree $p+1$ in (7), i.e. $H_{p+1}$.

$$\dot{Z}^{(p)} = \Lambda \cdot Z^{(p)} + \sum_{j=2}^{p} D_j(Z^{(p)}) + \sum_{j=p+1}^{\infty} \left( D_j(Z^{(p)}) + C_j(Z^{(p)}) \right) \quad (5)$$

$$\dot{Z}^{(p+1)} = \Lambda \cdot Z^{(p+1)} + \sum_{j=2}^{p+1} D'_j(Z^{(p+1)}) + \sum_{j=p+2}^{\infty} \left( D'_j(Z^{(p+1)}) + C'_j(Z^{(p+1)}) \right) \quad (6)$$

$$Z^{(p)} = H_{p+1}(Z^{(p+1)}) = Z^{(p+1)} + h_{p+1}(Z^{(p+1)}) \quad (7)$$

**Proof of Lemma 1** Consider the transformation in (7), where $h_{p+1}$ is a column vector whose elements are the homogeneous polynomials of degree $p+1$ in $Z^{(p+1)}$. The $(2i-1)$-th and $2i$-th elements of $h_{p+1}$ are shown in (8).

$$\begin{cases} h_{p+1,2i-1} = \sum_{\alpha=1}^{N}\sum_{\beta=\alpha}^{N}\cdots\sum_{\gamma=\eta}^{N} h_{p+1,2i-1,\alpha\beta\cdots\eta\gamma} \underbrace{z_{\alpha}^{(p+1)}z_{\beta}^{(p+1)}\cdots z_{\gamma}^{(p+1)}}_{p+1 \text{ terms in total}} \\ h_{p+1,2i} = \bar{h}_{p+1,2i-1} \end{cases} \quad (8)$$

Substitute (7) and (8) into (5) and obtain a transformed multi-oscillator system as shown in (9).

$$\begin{aligned} \dot{z}_{2i-1}^{(p+1)} &= \lambda_{2i-1} z_{2i-1}^{(p+1)} + \sum_{\alpha=1}^{N}\sum_{\beta=\alpha}^{N} \mu_{2i-1,\alpha\beta} z_{\alpha}^{(p+1)} z_{\beta}^{(p+1)} + \cdots \\ &\quad + \sum_{\alpha=1}^{N}\sum_{\beta=\alpha}^{N}\cdots\sum_{\eta=\chi}^{N} \mu_{2i-1,\alpha\beta\cdots\eta} \underbrace{z_{\alpha}^{(p+1)} z_{\beta}^{(p+1)}\cdots z_{\eta}^{(p+1)}}_{p \text{ terms in total}} \\ &\quad + \sum_{\alpha=1}^{N}\sum_{\beta=\alpha}^{N}\cdots\sum_{\gamma=\eta}^{N} c_{p+1,2i-1,\alpha\beta\cdots\eta\gamma} \underbrace{z_{\alpha}^{(p+1)} z_{\beta}^{(p+1)}\cdots z_{\gamma}^{(p+1)}}_{p+1 \text{ terms in total}} \\ &\quad + \cdots \\ \dot{z}_{2i}^{(p+1)} &= \bar{\dot{z}}_{2i-1}^{(p+1)} \\ i &= 1,\cdots,N/2 \end{aligned} \quad (9)$$

where

$$c_{p+1,2i-1,\alpha\beta\cdots\gamma} = \mu_{i,\mathrm{intra},\alpha\beta\cdots\gamma} + h_{p+1,2i-1,\alpha\beta\cdots\gamma} \cdot (\lambda_{2i-1} - \lambda_{\alpha} - \cdots - \lambda_{\gamma}) \quad (10)$$

Let the coefficients of terms of degree $p+1$ in (9) follow (3), i.e. admitting (11), and then we can obtain (6).

$$\begin{cases} h_{p+1,i,\mathrm{inter},\alpha\beta\cdots\gamma} = \dfrac{c_{p+1,i,\mathrm{inter},\alpha\beta\cdots\gamma}}{\underbrace{\lambda_{\alpha} + \cdots + \lambda_{\gamma}}_{p+1 \text{ terms in total}} - \lambda_{i}} \\ h_{p+1,i,\mathrm{intra},\alpha\beta\cdots\gamma} = \dfrac{c_{p+1,i,\mathrm{intra},\alpha\beta\cdots\gamma} - \mu_{i,\mathrm{intra},\alpha\beta\cdots\gamma}}{\underbrace{\lambda_{\alpha} + \cdots + \lambda_{\gamma}}_{p+1 \text{ terms in total}} - \lambda_{i}} \end{cases} \quad (11)$$

Note that when using transformation in (7) to transform (5) into (9), calculating the inverse of the coefficient matrix, e.g. left-hand side of equation (6) of paper [37], is implicitly required. This coefficient matrix is actually a function of all state variables $Z_{p+1}$, which is a near-identity matrix when close to the origin. However, if system states are far away from the origin, that matrix may become singular such that (9) cannot be obtained any more. An upper bound of the validity limit is introduced in paper [37], which indicates that (9) can be obtained only when the system states are close to the origin. ∎

**Remark** Given the transformation (7) obtained in Lemma 1, the *validity region of the transformation*, denoted as $\Omega_{p+1}$, is defined as the set of system states $Z^{(p+1)}$ within a region, where any point in the set does not lead to a singular coefficient matrix.

**Proof of Theorem 1** Given a multi-oscillator system (1), its *modal space representation* can be obtained as (12) by the transformation in (13), where $Z^{(1)}$ is the vector of state variables in the modal space and $U$ is the matrix of right eigenvectors.

$$\dot{Z}^{(1)} = U^{-1}F(UZ^{(1)}) \quad (12)$$

$$X = U \cdot Z^{(1)} \quad (13)$$

Taylor expansion of (12) can be written as

$$\dot{Z}^{(1)} = \Lambda \cdot Z^{(1)} + \sum_{j=2}^{\infty}\left(D_{j}^{<1>}(Z^{(1)}) + C_{j}^{<1>}(Z^{(1)})\right) \quad (14)$$

Apply Lemma 1 with $p=1$, then we can transform (14) into

$$\dot{Z}^{(2)} = \Lambda \cdot Z^{(2)} + D_{2}^{<2>}(Z^{(2)}) + \sum_{j=3}^{\infty}\left(D_{j}^{<2>}(Z^{(2)}) + C_{j}^{<2>}(Z^{(2)})\right) \quad (15)$$

Apply Lemma 1 for $k-2$ times respectively with $p=2, \ldots, k-1$, then we can transform (15) into (16).

$$\dot{Z}^{(k)} = \Lambda \cdot Z^{(k)} + \sum_{j=2}^{k} D_{j}^{<k>}(Z^{(k)}) + \sum_{j=k+1}^{\infty}\left(D_{j}^{<k>}(Z^{(k)}) + C_{j}^{<k>}(Z^{(k)})\right) \quad (16)$$

When the order $k$ approaches infinity, the convergence has to be considered. Since investigating the convergence issue is not a trivial task [38] and it is not the focus of this paper, we assume the convergence of this process holds when $k$ approaches infinity. Then, (2) will be achieved eventually, i.e. $Z=Z^{(\infty)}$, and the transformation $H$ in (4) is composed by $H_1$ in (13) and $H_{p+1}$ in (7) with $p=1, 2, \ldots$. ∎

In practice, it is hard to deal with an infinite number of transformations. Still, for any expected order $k$, we can use the truncated system as an approximation for practical applications. The following gives three corollaries of the nonlinear modal decoupling for any expected order $k$ with the help of the $k$-jet concept. Then, the decoupled $k$-jet system is introduced.

**Definition 4 ($k$-jet equivalence [21])** Assume $F(X)$ and $G(X)$ are two vector functions of the same dimension. We say that $F(X)$ and $G(X)$ are $k$-jet equivalent at $X_0$, or $F(X)$ is a $k$-jet equivalence of $G(X)$ and vice versa, *iff* corresponding terms in the Taylor expansions of $F(X)$ and $G(X)$ at $X_0$ are identical up to order $k$.

Then, a $k$-jet system of (1) can be rewritten in (17). The errors between the solutions of these two systems (1) and (17) are totally due to the truncation of high-order terms, whose impact will be investigated in the case studies. The following will start from (17) and analyze the nonlinear modal decoupling.

$$\dot{X} = AX + \sum_{j=2}^{k} A_{j}(X) \quad (17)$$

where $A_j(X)$ is a vector function, each of whose elements is a weighted sum of all homogeneous polynomials of degree $j$ in $X$.

**Corollary 1 ($k$-th order nonlinear modal decoupling)** Given a multi-oscillator system in (17), if the resonance does not exist up to the given order $k$, then the $k$-th *order nonlinearly mode-decoupled system* can be achieved as (18) by the $k$-th *order decoupling transformation $H^{(k)}$* in (19).

$$\dot{Z}^{(k)} = \Lambda \cdot Z^{(k)} + \sum_{j=2}^{k} D_{j}(Z^{(k)}) + \sum_{j=k+1}^{\infty}\left(D_{j}(Z^{(k)}) + C_{j}(Z^{(k)})\right) \quad (18)$$

$$X = H^{(k)}(Z^{(k)}) = (H_1 \circ H_2 \circ \cdots \circ H_k)(Z^{(k)}) \quad (19)$$

where $Z^{(k)}$ is the vector of state variables in the $k$-th order mode-decoupled space, $D_j$ and $C_j$ are vector functions where each of their elements is a weighted sum of the terms of degree $j$ in $Z^{(k)}$. $D_j$ only contains intra-modal terms, while $C_j$ only contains inter-modal terms.

**Corollary 2.** The *validity region* for the transformation $H^{(k)}$, denoted as $\Omega^{(k)}$, is

$$\Omega^{(k)} = \bigcap_{p=2}^{k} \Omega_p \tag{20}$$

**Corollary 3.** The inverse coordinate transformation, i.e. the inverse of the transformation $H_{p+1}$ in (7), can be approximated by a power series

$$z_i^{(p+1)} = z_i^{(p)} + \sum_{\alpha=1}^{N}\sum_{\beta=\alpha}^{N} s_{i,\alpha\beta} z_\alpha^{(p)} z_\beta^{(p)} + \cdots$$
$$+ \sum_{\alpha=1}^{N}\sum_{\beta=\alpha}^{N}\cdots\sum_{\gamma=\eta}^{N} s_{i,\alpha\beta\cdots\gamma} \underbrace{z_\alpha^{(p)} z_\beta^{(p)} \cdots z_\gamma^{(p)}}_{m\text{ terms in total}} + \cdots \tag{21}$$

**Remark** Based on existing literature, it is difficult to obtain such an inverse transformation in an explicit form or even a reliably transformation of a single point from the coordinates of $Z^{(p)}$ to that of $Z^{(p+1)}$. It was reported that the effectiveness of solving the nonlinear algebraic equation (7) by a certain iterative algorithm with an initial guess of $Z^{(p)}$ largely depends on that initial guess. The iterative algorithm may either diverge or converge to a different point in $Z^{(p)}$ [39][40]. Actually, the inverse of (7) can be written as (22). An approximate analytical expression to (22) is provided by Corollary 3. The proof is omitted while the idea is quite straightforward: (i) First, assume that the inverse transformation (22) follows the polynomial form, as shown in (21) which is the $i$-th equation of (22) where the coefficient $s$ of each term is an unknown. Substitute (21) into the right side of (7) and equate both sides term by term in $Z^{(p)}$ to formulate equations in $s$. Solving those equations for $s$ and substituting them back to (21) will give the inverse transformation. Note that those formulated equations in $s$ can always be solved due to the characteristic of (7), i.e. the function $h$ only contains homogeneous polynomials of degree $p+1$ in $Z^{(p)}$ and the formulated equations can always be solved order by order from low to high.

$$Z^{(p+1)} = Z^{(p)} + h_{p+1}^{-1}(Z^{(p)}) \tag{22}$$

**Definition 5. (Decoupled $k$-jet system)** By ignoring terms with orders higher than $k$ in (18), we obtain a special $k$-jet system of (18), called a *decoupled $k$-jet system*.

$$\dot{Z}_{\text{jet}}^{(k)} = \Lambda \cdot Z_{\text{jet}}^{(k)} + \sum_{j=2}^{k} D_j(Z_{\text{jet}}^{(k)}) \tag{23}$$

**Remark** Generally speaking, the nonlinearities maintained in the decoupled systems (18) by the intra-modal terms, i.e. $D_j$, can follow any pre-designed form and then defines a corresponding $k$-th order nonlinear transformation $H^{(k)}$. If without any a priori knowledge about the nonlinear characteristics of the original system (17), there could be an infinite number of ways to intentionally design the intra-modal terms, such that the resulting decoupled $k$-jet system (23) by different ways will differ from each other in terms of the size and shape of their validity regions. Also note that the equations in (23) about one mode are completely independent with those about any other mode, while the nonlinearities up to order $k$ within each individual mode are still maintained. Next, two theorems about the decoupled $k$-jet are introduced.

**Theorem 2 (Real-valued decoupled $k$-jet).** The decoupled $k$-jet in (23) is equivalent to a real-valued system, called a real-valued decoupled $k$-jet.

**Proof of Theorem 2** Since there may be multiple ways to construct a real-valued decoupled $k$-jet, we only provide the construction leading to two coordinates respectively having physical meanings similar to displacement and velocity.

The differential equations for mode $i$ in the decoupled $k$-jet are shown in (24), which are the $(2i\text{-}1)$-th and $2i$-th equations of (24). Note that the two state variables in (24) are complex-valued and those $\mu$-coefficients are determined in (7).

$$\begin{cases} \dot{z}_{2i-1}^{(k)} = \lambda_{2i-1} z_{2i-1}^{(k)} + \sum_{\alpha=1}^{N}\sum_{\beta=\alpha}^{N} \mu_{2i-1,\alpha\beta} z_\alpha^{(k)} z_\beta^{(k)} + \cdots \\ \qquad + \sum_{\alpha=1}^{N}\sum_{\beta=\alpha}^{N}\cdots\sum_{\rho=\gamma}^{N} \mu_{2i-1,\alpha\beta\cdots\rho} \underbrace{z_\alpha^{(k)} z_\beta^{(k)} \cdots z_\rho^{(k)}}_{k\text{ terms in total}} \\ \dot{z}_{2i}^{(k)} = \overline{\dot{z}}_{2i-1}^{(k)} \end{cases} \tag{24}$$

Because the two state variables in (24) are a conjugate pair, the right-hand sides of the first and second equations can be denoted respectively as $a+jb$ and $a$-$jb$, where $a$ and $b$ are real-valued functions in $Z^{(k)}$, $\lambda$ and $\mu$. Then, apply the coordinate transformation in (25) and (26) to yield (27), where all parameters and variables are real-valued. ∎

$$\begin{bmatrix} z_{2i-1}^{(k)} \\ z_{2i}^{(k)} \end{bmatrix} = U_{\text{mode }i}^{-1} \begin{bmatrix} w_{2i-1} \\ w_{2i} \end{bmatrix} \tag{25}$$

$$U_{\text{mode }i} = \begin{bmatrix} \lambda_{2i-1} & \lambda_{2i} \\ 1 & 1 \end{bmatrix} \tag{26}$$

$$\begin{cases} \dot{w}_{2i-1} = \upsilon_{i10} w_{2i-1} + \sum_{l=1}^{k} \upsilon_{i0l} w_{2i}^l + \sum_{\substack{j\geq 1, l\geq 0 \\ (j,l)\neq(1,0)}}^{j+l\leq k} \upsilon_{ijl} w_{2i-1}^j w_{2i}^l \\ \dot{w}_{2i} = w_{2i-1} + \sum_{\substack{j\geq 0, l\geq 0}}^{2\leq j+l\leq k} \nu_{ijl} w_{2i-1}^j w_{2i}^l \end{cases} \tag{27}$$

**Remark** Note that the transformation in (13) also need to be normalized. The purpose of a normalization is to make the new coordinates of (27) have a scale comparable to that of (23). The normalization introduced in [31] is adopted here:

(i) Classify the elements in the left eigenvector (complex-valued) related to the displacement into two opposing groups based on their angles;
(ii) Calculate the sum of the coefficients in one group;
(iii) Divide the left eigenvector by that sum;
(iv) Do such normalization for all left eigenvectors.

**Theorem 3 (Error estimation)** Given the multi-oscillator system in (17), its $k$-th order nonlinearly mode-decoupled system follows (18) and the corresponding decoupled $k$-jet follows (23). Assume the convergence of the decoupling transformation [38] and also assume all eigenvalues $\lambda$ in matrix $\Lambda$ to satisfy Re{$\lambda$} < $\alpha$ < 0. Let $X(t)$, $Z^{(k)}(t)$ and $Z^{(k)}_{\text{jet}}(t)$ be the solutions of (17), (18) and (23), respectively, under an identical initial condition $X_0= Z^{(k)}_0=Z^{(k)}_{\text{jet}0}$. Then, there exists constants $\varepsilon_1>0$ and $c>0$ which are independent of $k$ such that for any $\varepsilon$ in $0 \le \varepsilon \le \varepsilon_1$, $|X(0)| \le \varepsilon$ implies

$$e_k\left(t, X_0\right) \triangleq \left\| X(t) - H^{(k)}\left(Z^{(k)}_{\text{jet}}(t)\right) \right\| \le c\varepsilon^{k+1}e^{-\alpha t/2} \qquad (28)$$

where $|\cdot|$ represents a type of norm. Note that this theorem indicates that a better convergency can be achieved for any $t$ when $k$ increases.

**Proof of Theorem 3** It is easy to see that the two systems in (17) and (18) are equivalent in $\Omega^{(k)}$, over the transformation $H^{(k)}$ in (19). To show the error $e_k(t,X_0)$ approaches zero, we only need to show that error defined in (29), or equivalently (30), approaches zero:

$$\hat{e}_k\left(t, X_0\right) \triangleq \left\| H\left(Z^{(k)}(t)\right) - H\left(Z^{(k)}_{\text{jet}}(t)\right) \right\| \qquad (29)$$

$$\tilde{e}_k\left(t, X_0\right) \triangleq \left\| Z^{(k)}(t) - Z^{(k)}_{\text{jet}}(t) \right\| \qquad (30)$$

The rest of the proof is omitted since it is similar to the Theorem 5.3.4 in [41]. ∎

Given the $k$-th order nonlinearly mode-decoupled system in (18) and the corresponding decoupled $k$-jet system in (24), the $k$-th order nonlinear modal decoupling is $H^{(k)}$ in (19), whose inverse is assumed to be $(H^{(k)})^{-1}$, which is the composition of inverse transformations in (22) with different $p$. Denote $X(t,X_0)$ as the solution of (23) under the initial condition $X_0$ and denote $Z^{(k)}_{\text{jet}}(t, Z^{(k)}_{\text{jet}0})$ as the solution of (24) under the initial condition $Z^{(k)}_{\text{jet}0}$ where $Z^{(k)}_{\text{jet}0}=(H^{(k)})^{-1}(X_0)$. For a certain given $\varepsilon>0$, the *validity region $\Omega_\varepsilon$ of the decoupled $k$-jet* system is defined in (31).

$$\Omega_\varepsilon = \left\{ X_0 \,\middle|\, \left\| X(t, X_0) - H^{(k)}\left(Z^{(k)}_{\text{jet}}(t, Z^{(k)}_{\text{jet}0})\right) \right\| < \varepsilon \right\} \qquad (31)$$

Note that the validity region $\Omega_\varepsilon$ relies on the selection of $\varepsilon$. A larger $\varepsilon$ will lead to a larger validity region.

## III. Nonlinear Modal Decoupling of Power Systems

This section will apply the proposed nonlinear modal decoupling analysis to power systems. Firstly, the nonlinear differential equations of a multi-machine power system is introduced. Then, two forms of desired modal nonlinearity for the decoupled $k$-jet system are proposed. Finally, the first-integral based method is applied to the decoupled $k$-jet systems for stability analysis.

### A. Power system model

An $m$-machine power system is modeled by (32) and (33), where each generator is represented by a 2$^{\text{nd}}$ order classic model:

$$\ddot{\delta}_i + \frac{D_i}{2H_i}\dot{\delta}_i + \frac{\omega_s}{2H_i}\left(P_{\text{m}i} - P_{\text{e}i}\right) = 0 \qquad (32)$$

$$P_{\text{e}i} = E_i^2 G_i + \sum_{j=1, j\neq i}^{m}\left(C_{ij}\sin(\delta_i - \delta_j) + D_{ij}\cos(\delta_i - \delta_j)\right) \qquad (33)$$

where $i \in \{1,2,\dots,m\}$, $\delta_i$, $P_{\text{m}i}$, $P_{\text{e}i}$, $E_i$, $H_i$ and $D_i$ respectively represent the absolute rotor angle, mechanical power, electrical power, electromotive force, the inertia constant and damping constant of machine $i$, and $G_i$, $C_{ij}$, and $D_{ij}$ represent network parameters including loads modeled by constant impedances. The system state vector $X$ has a dimension of $2m$.

$$X = [\delta_1, \dot{\delta}_1, \delta_2, \dot{\delta}_2, \cdots, \delta_m, \dot{\delta}_m]^T \qquad (34)$$

**Remark** The $m$-machine power system modeled by (32) and (33) has $m$-1 pairs of conjugate complex eigenvalues, which respectively correspond to $m$-1 oscillatory modes, and two real eigenvalues (including one zero eigenvalue) [45]. We will focus on $m$-1 oscillatory models, which mainly determine rotor angle dynamics and stability of the power system. Thus, after the first coordinate transformation towards $Z^{(1)}$, only $2m$-2 differential equations corresponding to those oscillatory modes are kept for further analysis, i.e. $N=2m$-2. Denote the eigenvalues of these oscillatory modes as $\lambda_1, \lambda_2, \dots, \lambda_{2m\text{-}2}$, where $\lambda_{2i\text{-}1}$ and $\lambda_{2i}$ ($i$=1,2,..., $m$-1) belong to one conjugate pair.

Next, we present two ways to choose the desired modal nonlinearity respectively under the SMIB assumption and another proposed small transfer (ST) assumption.

### B. Nonlinear modal decoupling with the SMIB assumption

**SMIB assumption** [31][42]-[44][46][47] The nonlinearity associated with each oscillatory mode has the same form as an SMIB system.

In practice, this assumption is widely made by scholars and engineers in power systems. For instance, a power system that consists of two areas being weakly interconnected is often simplified to an SMIB system for stability studies regarding the inter-area oscillation mode. In the following, we study a general multi-machine power system. The goal is to find a way to intentionally transform the nonlinear terms of each decoupled system into the form of an SMIB system.

**Desired decoupled system for mode $i$** The desired decoupled system about mode $i$, i.e. eigenvalues $\lambda_{2i\text{-}1}$ and $\lambda_{2i}$, can be written as (35) [48].

$$\ddot{y}_i + \alpha_i\dot{y}_i + \beta_i\left(\sin(y_i + y_{is}) - \sin y_{is}\right) = 0 \qquad (35)$$

where $y_i$ is the generalized angle coordinate of mode $i$, while $\alpha_i$, $\beta_i$ and $y_{is}$ are constants that can be uniquely determined by

$$\begin{cases} y_{is} = \sum_{j=1}^{m} \tau_{ij} \delta_{js} \\ \alpha_i = -2\,\mathrm{Re}\{\lambda_{2i-1}\} \\ \beta_i = \dfrac{\lambda_{2i-1}\lambda_{2i}}{\cos y_{is}} \end{cases} \tag{36}$$

where $\tau_{ij}$ is the $i$-th row $j$-th column element from the matrix of the left eigenvectors defined using the state matrix of the linearization of (35) [31] and $\delta_{js}$ is the steady-state value of $\delta_j$.

**Nonlinear modal decoupling transformation** Assume each complex-valued decoupled system to be

$$\begin{cases} \dot{z}_{2i-1} = \lambda_{2i-1} z_{2i-1} + \sum_{j=2}^{k} \sum_{l=0}^{j} \mu_{i,l,j-l} z_{2i-1}^{l} z_{2i}^{j-l} \\ \dot{z}_{2i} = \bar{z}_{2i-1} \end{cases} \tag{37}$$

Toward the real-valued desired form of the decoupled system in (35), $\mu$-coefficients of intra-modal terms have yet to be determined. Apply the following coordinate transformation to (37) to obtain (40).

$$\begin{bmatrix} z_{2i-1} \\ z_{2i} \end{bmatrix} = V_{\mathrm{mode}\,i} \begin{bmatrix} \dot{y}_i \\ y_i \end{bmatrix} \tag{38}$$

where $V_{\mathrm{mode}\,i} = \dfrac{2}{\lambda_{2i-1} - \lambda_{2i}} \begin{bmatrix} 1 & -\lambda_{2i} \\ -1 & \lambda_{2i-1} \end{bmatrix}$ (39)

$$\begin{cases} \dfrac{d\dot{y}_i}{dt} + \alpha_i \dot{y}_i + \sum_{n=1}^{k} r_{in} y_i^n = 0 \\ \dfrac{dy_i}{dt} = \dot{y}_i \end{cases} \tag{40}$$

Coefficient $r_{in}$ is determined by (41) to make (35) and (40) have identical nonlinearities up to the $k$-th order.

$$r_{in} = \frac{\beta_i \cos\left(y_{is} + \dfrac{(n-1)\pi}{2}\right)}{n!} \tag{41}$$

### C. Nonlinear modal decoupling with the ST assumption

We also propose the following alternative assumption for each desired mode-decoupled system and compare its result with that from the SMIB assumption,

**Small transfer (ST) assumption** Assume the second equation of (11), i.e. $h_{p+1,i,\mathrm{intra},\alpha\beta\cdots\gamma}$, to be zero.

**Nonlinear modal decoupling transformation** The desired modal nonlinearity, i.e. $\mu$-coefficients, is chosen based on

$$\mu_{i,\mathrm{intra},\alpha\beta\cdots\gamma} = c_{p+1,i,\mathrm{intra},\alpha\beta\cdots\gamma} \tag{42}$$

**Remark** The desired decoupled system with the ST assumption might not be available before finishing the entire nonlinear modal decoupling process. However, the implementation is quite convenient, since we just need to let the second equation in (11) be zero. The physical insight behind this ST assumption is that we want to limit the propagation of nonlinear terms to higher orders over a transformation in (7), which can be seen in the example below.

Consider a system of two first-order differential equations with polynomial nonlinearities up to the 2nd order in (43), which is a special case of (17) with $N=2$ and $k=2$.

$$\begin{cases} \dot{z}_1 = \lambda_1 z_1 + b_{1,11} z_1^2 + b_{1,12} z_1 z_2 + b_{1,22} z_2^2 \\ \dot{z}_2 = \lambda_2 z_2 + b_{2,11} z_1^2 + b_{2,12} z_1 z_2 + b_{2,22} z_2^2 \end{cases} \tag{43}$$

Note that (43) only gives one of the two differential equations on each mode, so $\lambda_1$ and $\lambda_2$ are in fact the eigenvalues on two different modes, not the conjugate pair on one mode. This is for only simplicity of description. The idea is also applicable to other values of $N$ and $k$.

Intra-modal terms and inter-modal terms for these two equations in (43) are respectively listed in (44) and (45).

$$\left\{ b_{1,11} z_1^2, b_{2,22} z_2^2 \right\} \tag{44}$$

$$\left\{ b_{1,12} z_1 z_2, b_{1,22} z_2^2, b_{2,11} z_1^2, b_{2,12} z_1 z_2 \right\} \tag{45}$$

Then, consider a coordinate transformation by a 2nd order polynomial in (46).

$$\begin{cases} z_1 = u_1 + h_{1,11} u_1^2 + h_{1,12} u_1 u_2 + h_{1,22} u_2^2 \\ z_2 = u_2 + h_{2,11} u_1^2 + h_{2,12} u_1 u_2 + h_{2,22} u_2^2 \end{cases} \tag{46}$$

Substitute (46) into (43) and obtain a new system about $u$, where intra-modal terms and inter-modal terms are similar to (44) and (45) but defined about $u$ instead of $z$. In the new system, utilize the first equation of (11) to find coefficients $h$ to cancel its inter-modal terms as shown in (47) and obtain (48) where $P$, $Q$, $R$ and $S$ are polynomial functions, $S$ satisfies (50), and the coefficients of the intra-modal terms $h_{1,11}$ and $h_{2,22}$, denoted by $h_{\mathrm{intra}}$, are yet to be determined.

$$h_{1,12} = \frac{b_{1,12}}{\lambda_2}, \; h_{1,22} = \frac{b_{1,22}}{2\lambda_2 - \lambda_1}, \; h_{2,12} = \frac{b_{2,12}}{\lambda_1}, \; h_{2,11} = \frac{b_{2,11}}{2\lambda_1 - \lambda_2} \tag{47}$$

$$\begin{cases} \dot{u}_1 = \lambda_1 u_1 + (b_{1,11} - h_{1,11}\lambda_1) u_1^2 + \sum_{i=3}^{\infty} \sum_{\substack{0 \le j_1, j_2 \le i \\ j_1 + j_2 = i}} T_{1ij_1 j_2} u_1^{j_1} u_2^{j_2} \\ \dot{u}_2 = \lambda_2 u_2 + (b_{2,22} - h_{2,22}\lambda_2) u_2^2 + \sum_{i=3}^{\infty} \sum_{\substack{0 \le j_1, j_2 \le i \\ j_1 + j_2 = i}} T_{2ij_1 j_2} u_1^{j_1} u_2^{j_2} \end{cases} \tag{48}$$

where

$$\begin{cases} T_{1ij_1 j_2} = \dfrac{P_{10ij_1 j_2}(\lambda) R_{00ij_1 j_2}(b)}{Q_{10ij_1 j_2}(\lambda)} + \dfrac{P_{11ij_1 j_2}(\lambda) R_{11ij_1 j_2}(b)}{Q_{11ij_1 j_2}(\lambda)} \cdot S_{1ij_1 j_2}(h_{\mathrm{intra}}) \\ T_{2ij_1 j_2} = \dfrac{P_{20ij_1 j_2}(\lambda) R_{20ij_1 j_2}(b)}{Q_{20ij_1 j_2}(\lambda)} + \dfrac{P_{21ij_1 j_2}(\lambda) R_{21ij_1 j_2}(b)}{Q_{21ij_1 j_2}(\lambda)} \cdot S_{2ij_1 j_2}(h_{\mathrm{intra}}) \end{cases} \tag{49}$$

$$S(0) = 0 \tag{50}$$

Note that the information corresponding to the 2nd order nonlinearities in system (43) spreads out to all higher order nonlinear terms in system (48) through the transformation in (46). Theoretically speaking, such a transfer of nonlinearities does not impact the process of the nonlinear modal decoupling if keeping all nonlinear terms up to the infinite order in (48). However, the implementation of the nonlinear modal decoupling in practice can only keep nonlinear terms up to a finite order, say $k$, such that it will always be desired to keep the nonlinearities transferred to as few higher-order terms as possible. For specific systems, it might be possible that there exists a way to determine those non-zero $h_{intra}$ which can guarantee a minimum transfer, e.g. (51) where $D$ is a set of points $(u_1, u_2)$ containing all concerned dynamics of system (48). In general, without any a priori knowledge about the system, it might be preferred to let $h_{intra}=0$ in order to limit the transfer of nonlinearities, which is called the ST assumption.

$$\min_{h_{intra}}\left\{\max_{(u_1,u_2)\in D}\left[\left(\sum_{i=3}^{\infty}\sum_{\substack{0\le j_1,j_2\le i\\ j_1+j_2=i}}T_{1ij_1j_2}u_1^{j_1}u_2^{j_2}\right)^2+\left(\sum_{i=3}^{\infty}\sum_{\substack{0\le j_1,j_2\le i\\ j_1+j_2=i}}T_{2ij_1j_2}u_1^{j_1}u_2^{j_2}\right)^2\right]\right\} \quad (51)$$

### D. Nonlinear modal decoupling based stability analysis

The following assumption is adopted to create an explicit transient energy function for stability analysis: (52) and (53) holds for coefficients in (27).

$$\begin{cases} \upsilon_{ijl}=0 & \text{for all } j\ge1, l\ge1, j+l\le k, (j,l)\ne(1,0)\\ \nu_{ijl}=0 & \text{for all } j\ge0, l\ge0, 2\le j+l\le k \end{cases} \quad (52)$$

$$\upsilon_{i10}=0 \quad (53)$$

Then, (27) becomes (54) with the above assumption.

$$\begin{cases} \dot{w}_{2i-1}=\sum_{j=1}^{k}\upsilon_{ij}w_{2i}^{j}\\ \dot{w}_{2i}=w_{2i-1} \end{cases} \quad (54)$$

With the assumption in (52) and (53), a transient energy function of the real-valued decoupled $k$-jet about mode $i$, shown in (27), is

$$V_i(w_{2i-1},w_{2i})=\frac{w_{2i-1}^2}{2}+\int_0^{z_{2i}}\sum_{j=1}^{k}\upsilon_{ij}s^jds=\frac{w_{2i-1}^2}{2}+\sum_{j=1}^{k}\frac{\upsilon_{ij}}{j+1}w_{2i}^{j+1} \quad (55)$$

Given the system in (54) and its energy function in (55), the unstable equilibrium point (UEP) around the origin of (54) can be obtained by letting the right hand side of (54) be zero and solving the resulting algebraic equations. Denote the smallest positive real solution by $w_{2i,UEP}$. Here is Theorem 4:

**Theorem 4 (Stability criterion)** Given an initial condition of (54), e.g. $(w_{2i-1}(0), w_{2i}(0))$, if $V_i(0, w_{2i,UEP})$ is greater than $V_i(w_{2i-1}(0), w_{2i}(0))$, then the system (54) is stable; otherwise, the system (54) is unstable.

## IV. NUMERICAL STUDY

This section will present the numerical studies of the proposed nonlinear modal decoupling on two test power systems: the IEEE 3-machine, 9-bus system [49] and the New

England 10-machine, 39-bus system [50]. Each system is modeled by (32) and (33).

In the IEEE 9-bus power system, the detailed results from the proposed nonlinear modal decoupling will be presented: 1) two sets of decoupled system equations are respectively derived under the SMIB and ST assumptions; 2) numerical simulation results on the decoupled systems are created and compared to that from the normal form method; 3) stability on the original system is analyzed by means of analysis on decoupled systems. The New England 39-bus power system is then used to demonstrate the applicability of the proposed nonlinear modal decoupling method on a high-dimensional dynamical system.

### A. Test on the IEEE 9-bus system

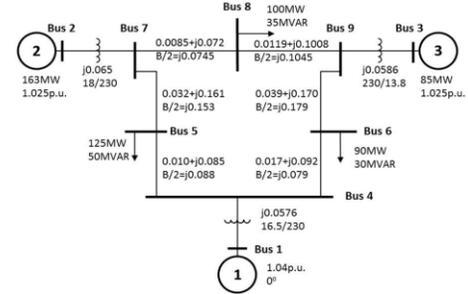

Fig. 1. IEEE 3-machine, 9-bus power system.

The following disturbance is considered: a temporary three-phase fault is added on bus 5 and cleared by tripping line 5-7 after a fault duration time. The critical clearing time (CCT) of this disturbance, i.e. the longest fault duration without causing instability, is found to be 0.17s. The post-disturbance system is represented by differential equations in (56) and has a stable equilibrium $x_{sep}$ = (3.12, 0, 3.12, 0, 3.12, 0), which is not the origin but is normal for a stable power system that has all generators operate coherently at one common speed after the disturbance. A 3rd order Taylor expansion of (56) gives an estimate of CCT equal to 0.16s, which has been very close to the accurate 0.17s, so the 3rd order Taylor expansion can credibly keep the stability information of the original system and is used below as the basis for deriving decoupled systems as well as the benchmark for comparison.

$$\begin{cases} \dot{x}_1=x_2+3.12\\ \dot{x}_2=-0.5x_2-1.14\cos(x_{13}-0.728)-6.25\sin(x_{13}-0.728)\\ \qquad -1.56\cos(x_{15}-0.463)-9.11\sin(x_{15}-0.463)-5.98\\ \dot{x}_3=x_4+3.12\\ \dot{x}_4=-0.5x_4-4.22\cos(x_{13}-0.728)+23.1\sin(x_{13}-0.728)\\ \qquad -6.04\cos(x_{35}+0.265)-38.0\sin(x_{35}+0.265)-5.98\\ \dot{x}_5=x_6+3.12\\ \dot{x}_6=-0.5x_6-12.3\cos(x_{15}-0.463)+71.6\sin(x_{15}-0.463)\\ \qquad -12.8\cos(x_{35}+0.265)+80.7\sin(x_{35}+0.265)-5.98\\ \qquad\qquad\qquad \text{where } x_{ij}=x_i-x_j \end{cases} \quad (56)$$

Following the nonlinear modal decoupling with the SMIB assumption and the ST assumption, the complex-valued decoupled 3-jet systems are respectively shown in (57) and (58), where the transformations are omitted. As a comparison, the counterpart from the $3^{rd}$ order normal form gives (59).

$$\begin{cases}
\dot{z}_1^{(3)} = (-0.25 + j12.9)z_1^{(3)} - j2.83z_1^{(3)}z_{(3)2} \\
\quad - j1.08\left(z_1^{(3)}\right)^3 - j1.42\left(z_2^{(3)}\right)^2 - j1.08\left(z_2^{(3)}\right)^3 - j1.42\left(z_2^{(3)}\right)^2 \\
\quad - j3.23\left(z_1^{(3)}\right)^2 z_2^{(3)} - j3.23z_1^{(3)}\left(z_2^{(3)}\right)^2 + \mathrm{O}\left(z^{(3)}\right)^4 \\
z_2^{(3)} = \overline{z}_1^{(3)} \\
\dot{z}_3^{(3)} = (-0.25 + j6.08)z_3^{(3)} - j1.52\left(z_3^{(3)}\right)^2 z_4^{(3)} \\
\quad - j0.51\left(z_3^{(3)}\right)^3 - j0.86\left(z_3^{(3)}\right)^2 - j0.51\left(z_4^{(3)}\right)^3 - j0.86\left(z_4^{(3)}\right)^2 \\
\quad - j1.72z_3^{(3)}z_4^{(3)} - j1.52z_3^{(3)}\left(z_4^{(3)}\right)^2 + \mathrm{O}\left(z^{(3)}\right)^4 \\
\dot{z}_4^{(3)} = \overline{z}_3^{(3)}
\end{cases} \tag{57}$$

$$\begin{cases}
\dot{z}_1^{(3)} = (-0.25 + j12.9)z_1^{(3)} + \mathrm{O}\left(z^{(3)}\right)^4 \\
\quad -(0.0019 + j0.0975)\left(z_1^{(3)}\right)^2 + (0.0057 - j0.097)\left(z_2^{(3)}\right)^2 \\
\quad +(0.0038 - j0.195)z_1^{(3)}z_2^{(3)} \\
\quad -(0.0144 + j0.372)\left(z_1^{(3)}\right)^3 - (1.3e-4 + j0.99)\left(z_1^{(3)}\right)^2 z_2^{(3)} \\
\quad +(0.048 - j1.11)z_1^{(3)}\left(z_2^{(3)}\right)^2 + (0.02 - j0.262)\left(z_2^{(3)}\right)^3 \\
z_2^{(3)} = \overline{z}_1^{(3)} \\
\dot{z}_3^{(3)} = (-0.25 + j6.08)z_3^{(3)} + \mathrm{O}\left(z^{(3)}\right)^4 \\
\quad -(0.023 + j0.57)\left(z_3^{(3)}\right)^2 + (0.07 - j0.566)\left(z_4^{(3)}\right)^2 \\
\quad +(0.047 - j1.14)z_3^{(3)}z_4^{(3)} \\
\quad -(0.008 + j0.098)\left(z_3^{(3)}\right)^3 + (0.002 - j0.365)\left(z_3^{(3)}\right)^2 z_4^{(3)} \\
\quad +(0.025 - j0.293)z_3^{(3)}\left(z_4^{(3)}\right)^2 + (0.017 - j0.092)\left(z_4^{(3)}\right)^3 \\
\dot{z}_4^{(3)} = \overline{z}_3^{(3)}
\end{cases} \tag{58}$$

$$\begin{cases}
\dot{z}_1^{(3)} = (-0.25 + j12.9)z_1^{(3)} + \mathrm{O}\left(z^{(3)}\right)^4 \\
\dot{z}_2^{(3)} = (-0.25 - j12.9)z_2^{(3)} + \mathrm{O}\left(z^{(3)}\right)^4 \\
\dot{z}_3^{(3)} = (-0.25 + j6.08)z_3^{(3)} + \mathrm{O}\left(z^{(3)}\right)^4 \\
\dot{z}_4^{(3)} = (-0.25 - j6.08)z_4^{(3)} + \mathrm{O}\left(z^{(3)}\right)^4
\end{cases} \tag{59}$$

Systems (57), (58) and (59) are respectively named ND-SMIB, ND-ST and NF. Their dynamical performances are compared with the same initial conditions under the post-disturbance condition. The error of each simulated system response is calculated compared to the "true" system response, which is simulated from the $3^{rd}$ Taylor expansion (56).

The errors $e(t)$ of these trajectories in the time domain are calculated by (29) and shown in Table I for four disturbances with increasing fault duration times from 0.01s to 0.15s. The last one gives a marginally stable case. The simulated trajectories from these systems and their time domain errors are shown in Fig. 2 to Fig. 5. From those figures and Table I, the ND-ST has the smallest error and the ND-SMIB has the largest error.

Then, the stability of the original system is studied using the ND-ST system (58). Transform (58) into real-valued equations by (25) to give (60).

TABLE I
TIME DOMAIN ERRORS OF SIMULATED RESPONSES

| FD | ND-SMIB | | ND-ST | | NF | |
|---|---|---|---|---|---|---|
| | $\mathrm{E}[e(t)]$ | $\mathrm{Std}[e(t)]$ | $\mathrm{E}[e(t)]$ | $\mathrm{Std}[e(t)]$ | $\mathrm{E}[e(t)]$ | $\mathrm{Std}[e(t)]$ |
| 0.01s | 0.95 | 1.06 | 0.07 | 0.08 | 0.17 | 0.18 |
| 0.05s | 1.19 | 1.36 | 0.12 | 0.14 | 0.44 | 0.50 |
| 0.10s | 3.93 | 4.36 | 0.40 | 0.46 | 2.41 | 2.74 |
| 0.15s | 15.31 | 14.01 | 1.82 | 2.21 | 16.47 | 18.67 |

$\mathrm{E}[e(t)]$ and $\mathrm{Std}[e(t)]$ are the expectation and standard deviation of the error signal $e(t)$, which is in degrees.

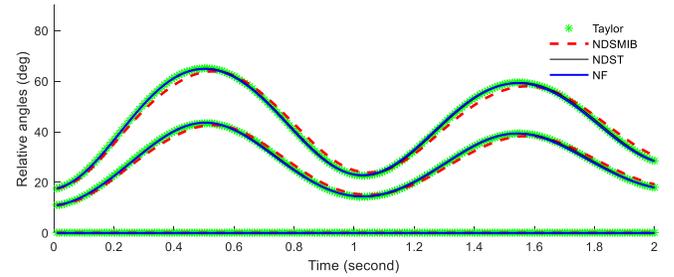

Fig. 2. Simulated system responses under the disturbance (fault duration = 0.01s) respectively by 3rd-order Taylor expansion, NDSMIB, NDST and NF.

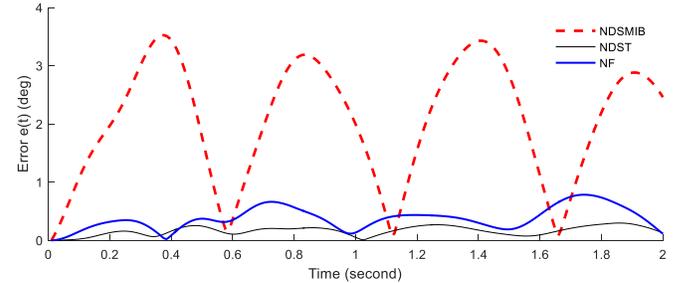

Fig. 3. Time domain error of the simulated system responses under the disturbance (fault duration = 0.01s) respectively by NDSMIB, NDST and NF.

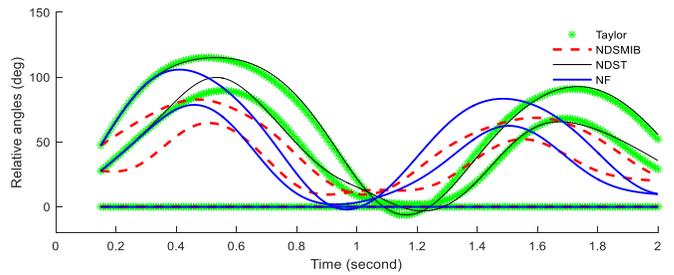

Fig. 4. Simulated system responses under the disturbance (fault duration = 0.15s) respectively by 3rd-order Taylor expansion, NDSMIB, NDST and NF.

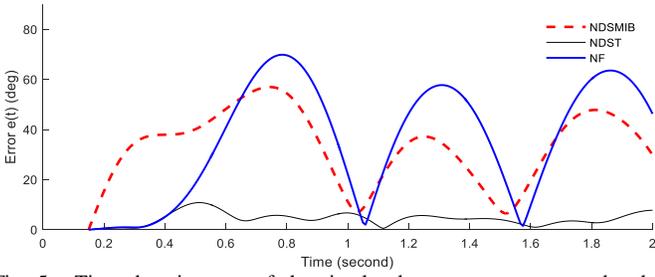

Fig. 5. Time domain error of the simulated system responses under the disturbance (fault duration = 0.15s) respectively by NDSMIB, NDST and NF.

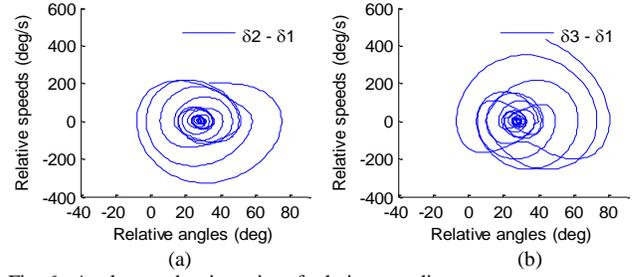

Fig. 6. Angle-speed trajectories of relative coordinates.

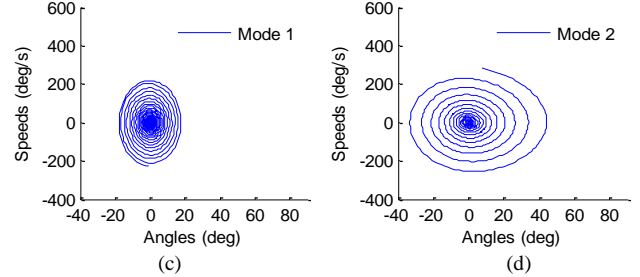

Fig. 7. Angle-speed trajectories in decoupled systems.

$$\begin{cases} \dot{w}_1 = -0.5w_1 - 166.0w_2 + 5.0w_2^2 + 35.3w_2^3 + 0.015w_1w_2 \\ \quad -0.139w_1w_2^2 + 0.016w_1^2w_2 + (1e-5) \cdot w_1^2 - (1e-4) \cdot w_1^3 \\ \dot{w}_2 = w_1 + (2e-8) \cdot w_1^2 + 0.008w_2^2 - (3e-5)w_1w_2 \\ \quad + (1e-4) \cdot w_1^3 + 0.05w_2^3 + (3e-4)w_1^2w_2 - 0.016w_1w_2^2 \\ \dot{w}_3 = -0.5w_3 - 37.1w_4 + 13.8w_4^2 + 5.13w_4^3 - 0.186w_3w_4 \\ \quad -0.087w_3w_4^2 + 0.015w_3^2w_4 + (6e-4) \cdot w_3^2 + (1e-4) \cdot w_3^3 \\ \dot{w}_4 = w_3 + (4e-6) \cdot w_3^2 + 0.093w_4^2 - 0.0013w_3w_4 \\ \quad -(3e-4) \cdot w_3^3 + 0.03w_4^3 - (2e-4)w_3^2w_4 - 0.015w_3w_4^2 \end{cases} \quad (60)$$

Simplify (60) to (61) using assumption in (52) and (53):

$$\begin{cases} \dot{w}_1 = -166.0w_2 + 5.0w_2^2 + 35.3w_2^3 \\ \dot{w}_2 = w_1 \\ \dot{w}_3 = -37.1w_4 + 13.8w_4^2 + 5.13w_4^3 \\ \dot{w}_4 = w_3 \end{cases} \quad (61)$$

Compare (61) with (60), the terms ignored according to (53) are actually either small or related to the damping effects. Thus, the stability analysis results on (61) may be conservative for systems in (60). Then, the first-integral based energy functions for the two modes are calculated to be (62).

$$\begin{cases} V_1(w_1, w_2) = \dfrac{w_1^2}{2} - 83w_2^2 + 1.6667w_2^3 + 8.825w_2^4 \\ V_2(w_3, w_4) = \dfrac{w_3^2}{2} - 18.55w_4^2 + 4.6w_4^3 + 1.2825w_4^4 \end{cases} \quad (62)$$

Let the right hand side of (61) be zeros and solve for the UEPs and get $w_{2,\text{UEP}} = 2.0986$ and $w_{4,\text{UEP}} = 1.6618$. The critical energy for the two modes are $V_1(0, w_{2,\text{UEP}}) = 178.9041$ and $V_2(0, w_{4,\text{UEP}}) = 20.3363$. Under different fault durations, the initial energy of the decoupled systems is shown in Table II, which tells that the initial energy of the system corresponding to the second mode first exceeds its critical energy when the fault duration reaches 0.16s while the initial energy corresponding to the first mode is always much smaller than its critical energy. Table II also shows that the CCT found by this analysis is 0.15s, which is fairly accurate when compared to 0.16s, the "true" CCT of the 3rd order Taylor expansion system.

TABLE II
Initial Energy of NDST Systems Under Different Fault Durations

| Fault Duration (s) | $V_1(w_1(0), w_2(0))$ | $V_2(w_3(0), w_4(0))$ |
|---|---|---|
| 0.01 | 0.0022 | 3.2072 |
| 0.05 | 0.0128 | 4.6495 |
| 0.10 | 0.1173 | 9.6173 |
| 0.15 | 0.6785 | 19.402 |
| 0.16 | 0.8643 | 22.092 |

Another benefit of the nonlinear modal decoupling is that the trajectory of each decoupled system can be drawn in the corresponding coordinates as a trajectory only about one mode. In that sense, the original system's trajectories regarding different modes are also nonlinearly decoupled. For the marginally stable case with fault duration =0.15s, Fig. 6 plots the trajectories of the original system in different coordinates while Fig. 7 visualizes the modal trajectories in the coordinates about each decoupled system. In this case, both oscillatory modes of the system are excited, so the original trajectories are tangled. However, the trajectory on each decoupled system is clean and easier to analyze.

### B. Test on the New England 39-bus system

This subsection will test the proposed nonlinear modal decoupling on the New England 10-machine, 39-bus power system [50]. Using the 2nd order Taylor expansion of the 20 nonlinear differential equations and the two assumptions in Section III, two sets of decoupled 2-jets can be obtained. A three-phase fault is added on bus 16 and cleared after 0.2 second by tripping the line 15-16. With the same initial condition under this fault, the 2nd order Taylor expansion of the original system, the two decoupled 2-jets, and the 2nd order normal form are simulated and compared in the original space, as shown in Fig. 9 and Fig. 10. Similar to the case study on the IEEE 9-bus system, the error of ND-ST is the smallest among the three.

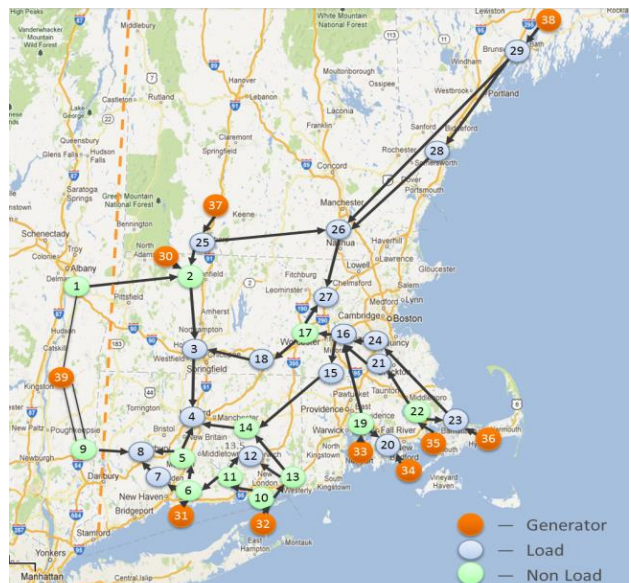

Fig. 8.  New England 39-bus power system.

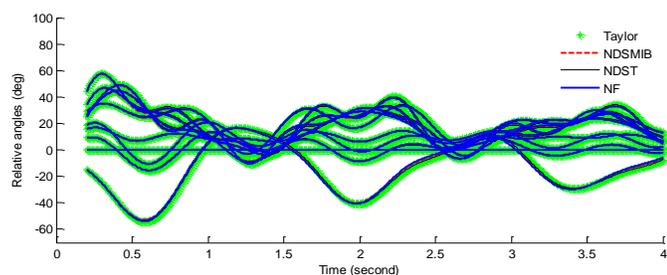

Fig. 9.  Simulated system responses under the disturbance (fault duration = 0.2s) respectively by 2nd-order Taylor expansion, NDST, NF and NFSMIB.

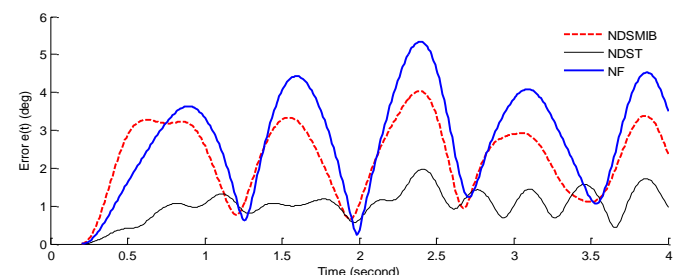

Fig. 10.  Time domain error of the simulated system responses under the disturbance (fault duration = 0.2s) respectively by ND-ST, NF and NF-SMIB.

## V. CONCLUSION

This paper proposes the nonlinear modal decoupling analysis to transform a general multi-oscillator system into a set of decoupled 2$^{nd}$ order single oscillator systems with polynomial nonlinearities up to a given order. Since the decoupled systems are low dimensional and independent with each other up to the given order, they can be easier analyzed compared to the original system, and they provide more physical insights to the dynamics of the original system. The analysis on each nonlinear mode-decoupled system makes all available techniques applicable to low dimensional nonlinear systems be also applicable to the original multi-oscillator system.

The derivation of the nonlinear modal decoupling adopts the idea of the normal form method, and the decoupling transformation turns out to be the composition of a set of nonlinear homogeneous polynomial transformations. The key step in deriving the nonlinear modal decoupling is the elimination of the inter-modal terms and the retention of nonlinearities only related to the intra-modal terms. The elimination of inter-modal terms can be achieved uniquely, while the intra-modal terms could be maintained in an infinite number of ways such that a desired form has to be specified.

Then, the nonlinear modal decoupling analysis is applied to power systems toward two forms of decoupled systems: (i) the single-machine-infinite-bus (SMIB) assumption; (ii) the small transfer (ST) assumption. Note that the ST assumption does not limit mode-decoupled systems to the power system models; rather, they can be any other type of oscillator systems if a priori knowledge or preference on the form of mode-decoupled systems is not available. Numerical studies on both a small IEEE 3-machine, 9-bus system, and a larger New England 10-machine, 39-bus system, show that the decoupled system under the ST assumption has a larger validity region than the decoupled systems under the SMIB assumption and the transformed linear system from the normal form method. It is also demonstrated that the decoupled systems can enable easier and fairly accurate analyses, e.g. on stability of the original system.

Future work includes calculation of validity region $\Omega^{(k)}$, finding the best way to maintain the intra-modal terms to achieve the largest validity region, the stability analysis and controller design based on the nonlinear modal decoupling.

## REFERENCES

[1]  S.Z. Cardon, A.S. Iberall, "Oscillations in biological systems", *Biosystems*, vol.3, no.3, pp.237-249, 1970.

[2]  G. Rogers, *Power System Oscillations*. New York: Springer, 2000.

[3]  Y. Kuramoto, *Chemical oscillations, waves, and Turbulence*, New York: Springer-Verlag, 1984.

[4]  A. Pluchino, V. Latora, A. Rapisarda, "Changing opinions in a changing world: a new perspective in sociophysics", *Int. J. Mod. Phys.*, vol.16, no.4, 2005.

[5]  B. V. Chirikov, "A universal instability of many-dimensional oscillator systems", *Physics Reports*, vol.52, no.5, pp.263-379, 1979.

[6]  S. Wiggins, *Introduction to applied nonlinear dynamical systems and chaos*. New York: Springer-Verlag, 1990.

[7]  M. A. Porter, J. P. Gleeson, *Dynamical Systems on Networks: A Tutorial*. Switzerland: Springer International Publishing, 2016.

[8]  A. H. Nayfeh, *Nonlinear oscillations*. John Wiley & Sons, 2008.

[9]  S. Boccaletti, V. Latora, Y. Moreno, M. Chavez, D. U. Hwang, "Complex networks: structure and dynamics", *Physics Reports*, vol.424, no.4-5, pp.175-308, 2006.

[10]  F. Dörfler and F. Bullo, "Synchronization in complex oscillator networks: A survey", *Automatica*, vol.50, no.6, pp.1539-1564, 2014.

[11]  A. Arenas, A. D. Guilera, J. Kurths, Y. Moreno, C. Zhou, "Synchronization in complex networks", *arXiv:0805.2976*, 2008

[12]  F. Dörfler, M. Chertkov and F. Bullo, "Synchronization in complex oscillator networks and smart grids", *Proc. Nat. Acad. Sci. USA*, vol.110, no.6, pp.2005-2010, Feb. 2013.

[13]  F. Dörfler, M. Chertkov and F. Bullo, "Synchronization assessment in power networks and coupled oscillators," *IEEE Conference on Decision and Control*, Maui, HI, 2012, pp. 4998-5003.

[14]  L. Glass, "Synchronization and rhythmic processes in physiology", *Nature*, doi:10.1038/35065745, 2001.

[15]  Z. Bai, "Krylove subspace techniques for reduced-order modeling of large-scale dynamical systems," *Applied Numerical Mathematics*, vol. 43, no. 1-2, pp. 9-44, Oct. 2002.

[16]  T. K. Caughey, M. J. O'Kelly, "Classical normal modes in damped linear dynamic systems", *ASME. J. Appl. Mech.*, vol.32, no.3, pp.583-588, 1965

[17]  E.T. Whittaker, *A Treatise on the Analytical Dynamics of Particles and Rigid Bodies*, 4th Edition. Cambridge: Cambridge University Press, 1985

[18]  M.T. Chu, N. Del Buono, "Total decoupling of general quadratic pencils, part I: theory," *J. Sound Vibration*, vol.309, no.1-2, pp.96-111, 2008

[19] M.T. Chu, N. Del Buono, "Total decoupling of general quadratic pencils, part II: structure preserving isospectral flows," *J. Sound Vibration*, vol.309, no.1-2, pp.112-128, 2008

[20] F. Ma, M. Morzfeld, A. Imam, "The decoupling of damped linear systems in free or forced vibration," *J. Sound Vibration*, vol.329, no.15, pp.3182-3202, 2010

[21] F. J. Doyle, F. Allgower and M. Morari, "A normal form approach to approximate input-output linearization for maximum phase nonlinear SISO systems," *IEEE Trans. Autom. Control*, vol. 41, no. 2, pp. 305-309, Feb 1996.

[22] A. Ilchmann and M. Muller, "Time-Varying Linear Systems: Relative Degree and Normal Form," *IEEE Trans. Autom. Control*, vol. 52, no. 5, pp. 840-851, May 2007.

[23] L. B. Freidovich and H. K. Khalil, "Performance Recovery of Feedback-Linearization-Based Designs," *IEEE Trans. Autom. Control*, vol. 53, no. 10, pp. 2324-2334, Nov. 2008.

[24] W. M. Wonham and A. S. Morse, "Decoupling and pole assignment in linear multivariable systems: A geometric approach," *SIAM J. Contr.*, vol. 8, no. 1, pp. 1-18, Feb. 1970.

[25] L.O. Chua, H. Kokubu, "Normal forms for nonlinear vector fields. I. Theory and algorithm," *IEEE Trans. Circuits Syst.*, vol.35, no.7, pp.863-880, Jul 1988

[26] F. Takens, Singularities of vector fields, *Publ. Math. I.H.E.S.*, vol.43, pp.47-100, 1974

[27] S. Ushiki, Normal forms for singularities of vector fields, *Japan J. Appl. Math.*, vol.1, pp.1-37, 1984

[28] F. Howell and V. Venkatasubramanian, "A modal decomposition of the Hopf normal form coefficient," *IEEE Trans. Autom. Control*, vol. 46, no. 7, pp. 1080-1083, Jul 2001.

[29] W. Kang and A. J. Krener, "Normal forms of nonlinear control systems," *Chaos in automatic control*, pp. 345-376, CRC Press, 2006.

[30] G. Ledwich, "Decoupling for improved Modal Estimation," *IEEE PES General Meeting*, Tampa, FL, 2007

[31] B. Wang, K. Sun and X. Su, "A decoupling based direct method for power system transient stability analysis," *IEEE PES General Meeting*, Denver, CO, 2015

[32] V. Vittal, W. Kliemann, S. K. Starrett, A. A. Fouad, "Analysis of stressed power systems using normal forms," *IEEE International Symposium on Circuits and Systems*, San Diego, CA, 1992

[33] S. K. Starrett and A. A. Fouad, "Nonlinear measures of mode-machine participation [transmission system stability]," *IEEE Trans. Power Syst.*, vol. 13, no. 2, pp. 389-394, May 1998.

[34] H. Amano, T. Kumano, T. Inoue, "Nonlinear stability indexes of power swing oscillation using normal form analysis," *IEEE Trans. Power Syst.*, vol.21, no.2, pp.825-834, May 2006

[35] H. M. Shanechi, N. Pariz and E. Vaahedi, "General nonlinear modal representation of large scale power systems," *IEEE Trans. Power Syst.*, vol. 18, no. 3, pp. 1103-1109, Aug. 2003.

[36] V. I. Arnold, *Geometrical Methods in the Theory of Ordinary Differential Equations*. New York: Springer-Verlag, 1988.

[37] C. H. Lamarque, C. Touze, O. Thomax, "An upper bound for validity limits of asymptotic analytical approaches based on normal form theory," *Nonlinear Dyn.*, DOI:10.1007/s11071-012-0584-y.

[38] G. Cicogna, S. Walcher, "Convergence of normal form transformations: the role of symmetries," *Acta Applicandae Mathematicae*, vol. 70, no. 95, pp. 95-111, Jan. 2002.

[39] N. Kshatriya, U. Annakage, A.M. Gole and I.T. Fernando, "Improving the Accuracy of Normal Form Analysis," *IEEE Trans. Power Syst.*, vol. 20, no. 1, pp. 286-293, Feb. 2005

[40] E. Barocio, A.R. Messina, J. Arroyo, "Analysis of Factors Affecting Power System Normal Form Results", *Electric Power Systems Research*, vol.70, no.3, pp.223-236, 2004

[41] J. Murdock, *Normal Forms and Unfoldings for Local Dynamical Systems*. New York: Springer, 2003

[42] B. Wang, K. Sun, "Formulation and Characterization of Power System Electromechanical Oscillations," *IEEE Trans. Power Syst.*, DOI: 10.1109/TPWRS.2016.2535384

[43] B. Wang, K. Su, K. Sun, "Properties of the Frequency-Amplitude Curve," *IEEE Trans. Power Syst.*, DOI: 10.1109/TPWRS.2016.2553583

[44] N. Duan, B. Wang, K. Sun, "Analysis of Power System Oscillation Frequency Using Differential Groebner Basis and the Harmonic Balance Method," *IEEE PES General Meeting*, Denver, CO, 2015

[45] F. Saccomanno, "Electromechanical phenomena in a multimachine system," *Electric Power Systems*. New York: Wiley, 2003

[46] J. H. Chow, "Natural modes and their stability in power systems," *IEEE Conference on Decision and Control*, Fort Lauderdale, FL, USA, 1985, pp. 799-803.

[47] Y. Xue, T. Van Custem and M. Ribbens-Pavella, "Extended equal area criterion justifications, generalizations, applications," *IEEE Trans. Power Syst.*, vol. 4, no. 1, pp. 44-52, Feb 1989

[48] B. Wang, K. Sun, A. D. Rosso, E. Farantatos, N. Bhatt, "A study on fluctuations in electromechanical oscillation frequencies of power systems", *IEEE PES General Meeting*, National Harbor, 2014

[49] B. Wang and K. Sun, "Power system differential-algebraic equations," *arXiv preprint arXiv:1512.05185*, 2015.

[50] F. Hu, K. Sun, A.D. Rosso, E. Farantatos, N. Bhatt, "Measurement-based real-time voltage stability monitoring for load areas," *IEEE Trans. Power Syst.*, vol.31, no.4, pp.2787-2798, 2016

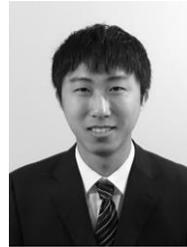
**Bin Wang** (S'14) received the B. S. and M.S. degrees in Electrical Engineering from Xi'an Jiaotong University, China, in 2011 and 2013, respectively. He is cur-rently pursuing the Ph.D. degree at the Department of EECS, University of Tennessee in Knoxville. His research interests include power system nonlinear dynamics, stability and control.

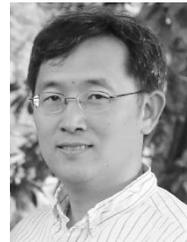
**Kai Sun** (M'06–SM'13) received the B.S. degree in automation in 1999 and the Ph.D. degree in control science and engineering in 2004 both from Tsinghua University, Beijing, China. He is currently an assistant professor at the Department of EECS, University of Tennessee in Knoxville. He was a project manager in grid operations and planning at the EPRI, Palo Alto, CA from 2007 to 2012. Dr. Sun is an editor of IEEE Transactions on Smart Grid and an associate editor of IET Generation, Transmission and Distribution. His research interests include power system dynamics, stability and control and complex systems.

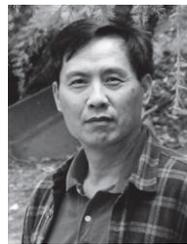
**Wei Kang** (M'91-F'08) received the B.S. and M.S. degrees from Nankai University, China, both in mathematics, in 1982 and 1985, respectively, and the Ph.D. degree in mathematics from the University of California, Davis, in 1991.

He is currently a Professor of applied mathematics at the U.S. Naval Postgraduate School, Monterey, CA. He was a visiting Assistant Professor of systems science and mathematics at Washington University, St. Louis, MO (1991-1994). He served as the Director of Business and International Collaborations at the American Institute of Mathematics (2008-2011). His research interest includes computational optimal control, nonlinear filtering, cooperative control of autonomous vehicles, industry applications of control theory, nonlinear $H_\infty$ control, and bifurcations and normal forms. His early research includes topics on Lie groups, Lie algebras, and differential geometry.

Dr. Kang is a fellow of IEEE. He was a plenary speaker in several international conferences of SIAM and IFAC. He served as an associate editor in several journals, including IEEE TAC and Automatica.